\documentclass{elsart}
\input{epsf}
\journal{Physica D}
\begin{document}

\begin{frontmatter}

\title{Nonlinear Analysis of the Eckhaus Instability:
Modulated Amplitude Waves and Phase Chaos with Non-zero Average Phase Gradient}

\author[ad1]{Lutz Brusch},
\author[ad2]{Alessandro Torcini} and
\author[ad1]{Markus B\"ar}
\address[ad1]{ Max-Planck-Institut f\"ur Physik
  komplexer Systeme, N\"othnitzer Stra{\ss}e 38, D-01187 Dresden, Germany}
\address[ad2]{Dipartimento di Energetica, Universit\'a di Firenze, 
via S. Marta 3 - I-50139 Firenze, Italy and
Istituto Nazionale di Ottica Applicata, L.go E. Fermi 6 - 
I-50125 Firenze, Italy}

\begin{abstract}

We analyze the Eckhaus instability of plane waves in the 
one-dimensional complex Ginzburg-Landau equation (CGLE) and 
describe the nonlinear effects arising 
in the Eckhaus unstable regime.
Modulated amplitude waves (MAWs) are quasi-periodic solutions of the CGLE 
that emerge near the Eckhaus instability of plane waves
and cease to exist due to saddle-node bifurcations (SN).
These MAWs can be characterized by their average phase gradient 
$\nu$ and by the spatial period $P$ of the periodic amplitude modulation. 
A numerical bifurcation analysis reveals the existence and stability 
properties of MAWs with arbitrary $\nu$ and $P$.
MAWs are found to be stable for large enough $\nu$ and intermediate
values of $P$. 
For different parameter values they are unstable to splitting and
attractive interaction between subsequent extrema of the amplitude. 
Defects form from perturbed plane waves for parameter values above the 
SN of the corresponding MAWs. 
The break-down of phase chaos with average phase gradient 
$\nu \ne 0$ (``wound-up phase chaos'') is thus related to these SNs.
A lower bound for the break-down of wound-up phase chaos
is given by the necessary presence 
of SNs and an upper bound by  the absence of the splitting
instability of MAWs.

\end{abstract}


\end{frontmatter}

\section{Introduction}
\label{intro}

The emergence of chaotic behaviour from ordered states in spatially
extended systems has been the subject of many recent experimental and
theoretical investigations  \cite{books,CH}.
Nonetheless, the mechanisms leading from stationary regimes to
chaotic (or spatially irregular) phases still pose many challenging
questions.
One of the most studied instabilities 
in extended oscillatory systems is the Eckhaus 
instability of plane waves \cite{eck}.

The occurrence of this instability has been experimentally 
observed in many quasi one-dimensional systems like
the oscillatory instability of a Rayleigh-B\'enard
convection pattern~\cite{janiaud},
hydrothermal waves \cite{h1,h2,garnier,CIAI4},
heated wire convection~\cite{wire},
sidewall convection~\cite{sidewall},
the Taylor-Dean system~\cite{dean} and
internal waves excited by the Marangoni effect~\cite{velarde}. 
The Eckhaus instability also plays an important role in 
the radial dynamics of spiral waves in the 
Belousov-Zhabotinsky reaction \cite{chem}.

\begin{figure}
\begin{center}
 \epsfxsize=0.7\hsize \mbox{\hspace*{-.06 \hsize} \epsffile{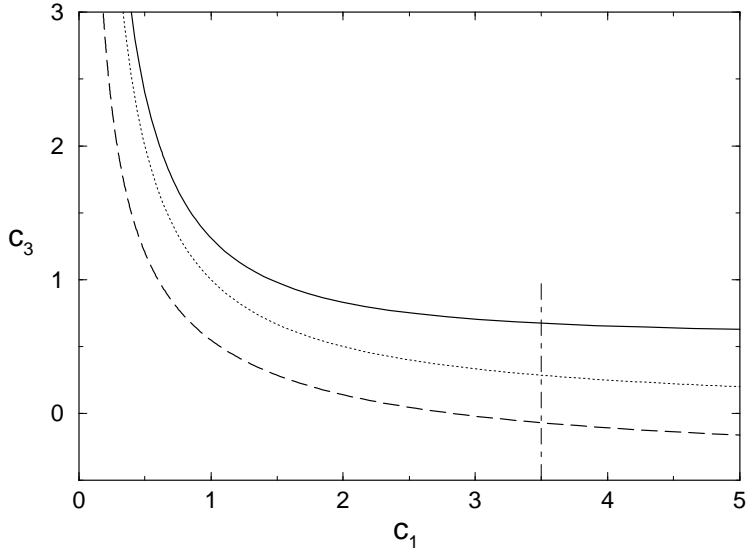} }
\end{center}
 \caption[Phase diagram of the CGLE]
 {Phase diagram of the one-dimensional CGLE.
 The dotted curve indicates the Benjamin-Feir-Newell line.
 Plane waves undergo an Eckhaus instability at values of $c_1, c_3$ 
 below this curve depending on their wavenumber.
 Above the dashed curve the Eckhaus instability is supercritical whereas it is
 subcritical below \cite{janiaud}.
 MAWs and wound-up phase chaos with $\nu>0$ can be observed between the 
 dashed and the full curve. 
 Defect chaos can occur only above the full curve \cite{MAW1,MAW2} which 
 denotes the saddle-node bifurcation of MAWs with $\nu=0$ and $P\to\infty$.
 The vertical dot-dashed line indicates
 the cut of the parameter space at $c_1=3.5$ studied in this paper.
 }
 \label{c1c3}
\end{figure}

The complex Ginzburg-Landau equation (CGLE) \cite{CH,review} is the 
 appropriate amplitude equation to describe the slow dynamics
near a supercritical transition to unidirectional traveling waves. 
In one spatial dimension, the CGLE  reads:
\begin{equation} 
  \partial_{t} A = A + (1+ ic_1) \partial_{x}^{2} A - (1-i c_3) |A|^2 A
  ~, \label{cgle}
\end{equation}
where $c_1$ and $c_3$ are real coefficients and the field 
$A=A(x,t)= |A(x,t)| {\rm e}^{i \varphi (x,t)}$ 
has complex values.  
As exact solutions the CGLE admits plane waves of the form
$ A_q (x,t) = a_q \mbox{e}^{i (q x - \omega_q t)} $, where
$q$ indicates the wavenumber, $a_q=\sqrt{1-q^2}$ and 
$ \omega_q = - c_3 + q^2(c_1+c_3) $.

A linear stability analysis \cite{EckDiP} of these solutions can 
be performed by considering the perturbed solution
$ \tilde{A}_q (x,t) = (a_q+\delta a) \mbox{e}^{i (q x - \omega_q t)} $, where
$\delta a \propto \mbox{e}^{i k x} \mbox{e}^{\sigma(k) t}$.
The growth rates associated to the complex perturbation $\delta a$ is
\begin{eqnarray} 
\sigma(k) &=& -k^2 -2 i q c_1 k - (1-q^2) \nonumber  \\
         &&   \pm \sqrt{(1+c_3^2)(1-q^2)^2 -[c_1 k^2-2 i q k-c_3 (1-q^2)]^2}~.
\label{sigmak}
\end{eqnarray}
The plane waves become linearly unstable to 
long wavelength perturbations ($k\to 0$) for
$q = q_E \equiv \sqrt{(1-c_1 c_3)/(2 (1+c_3^2) + 1-c_1 c_3)}$.
This limit is called Eckhaus instability \cite{eck}.
Above the Benjamin-Feir-Newell (BFN) line $1-c_1 c_3 = 0$, 
all plane waves are unstable to homogeneous perturbations.
For a given $q \ge q_E$,
the corresponding plane wave is linearly unstable 
against perturbations with wavenumbers $k$ inside 
the interval $0 < |k| < k_c$. $k_{c}$ increases
for increasing values of the parameters $c_1$ and $c_3$ \cite{notevHB}.

As noticed in \cite{EckCI1,EckCI2}, the Eckhaus instability
is a convective instability. 
Thus, it is relevant in the systems with periodic boundary conditions
considered here,
while it would be suppressed in fixed boundary conditions , {\it
e. g.} zero-flux or Dirichlet boundary conditions. 
In the latter geometries, the absolute instability of the plane 
waves has to be computed. 
It occurs for sufficiently large $q$ or/and $c_1$,$c_3$ values 
inside the Eckhaus unstable range.

Another interesting aspect of the Eckhaus instability is found 
when the nonlinearities of the CGLE are taken into account. 
A weakly nonlinear analysis \cite{janiaud} revealed that for
\begin{equation}
c_1^2 (1-6 c_3^2) + c_1 (2 c_3^3 + 16 c_3) - (8+c_3^2) > 0
\label{Ecksuper}
\end{equation}
the Eckhaus instability becomes supercritical, {\it i.e.}
the instabilities are saturated and the emerging quasi-periodic
solutions ({\it resp.} modulated amplitude waves) 
coexist with the unstable plane waves.
\begin{figure}
\begin{center}
 \epsfxsize=0.7\hsize \mbox{\hspace*{-.06 \hsize} \epsffile{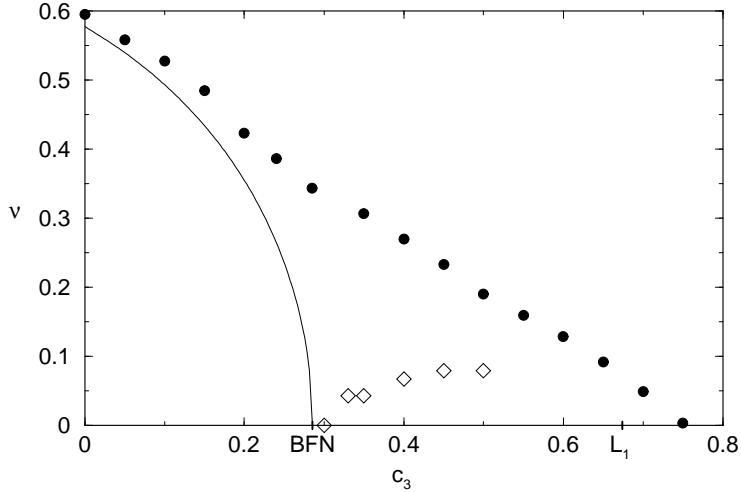} }
\end{center}
 \caption[Maximum conserved phase gradient $\nu_M (c_3)$]
 {Maximum (dots) conserved average phase gradient $\nu_M (c_3)$
for fixed $c_1=3.5$ 
obtained from numerical simulations (with system size $L=1,024-2,048$ and 
integration times $t\sim 10^5$) 
 for 50-70 different initial conditions 
(noise added to plane wave with wavenumber $q=\nu$). 
 For $\nu \le \nu_M$ no defects were present while above $\nu_M$ at least one
 initial condition caused defects. 
 The full curve denotes the Eckhaus instability of plane waves 
 that converges to the BFN line at $c_3=1/c_1$ for $\nu=0$.
 For $\nu$ above the diamonds regular states were observed after a transient 
 phase chaotic dynamics but below the diamonds most initial conditions led to persistent 
 spatio-temporal chaos \cite{torcini}.
L$_1$ denotes the lower bound for the occurrence of 
defect chaos in the thermodynamic limit as calculated in \cite{MAW1,MAW2}.
 }
 \label{FnuM}
\end{figure}
Numerical simulations 
\cite{janiaud,pumir,miguel,torcini,hager}
provided examples of such stable modulated amplitude waves (MAWs).
Stable MAWs have also been observed in experiments on 
surface-tension-driven hydrothermal waves \cite{h2} as well as 
on the Taylor-Dean system \cite{dean} and 
on internal waves excited by the Marangoni effect\cite{velarde}.

In addition, the CGLE exhibits two qualitatively
different spatiotemporal chaotic states known as phase chaos (when the 
modulus of the field $|A|$ is
bounded away from zero) and defect chaos (when the phase of $A$ displays
singularities where $|A|\!=\!0$) \cite{chao1,chate,saka,egolf}.
The subclass of MAWs with zero average phase gradient is important 
for understanding the transition from phase to defect chaos 
(see full curve in Fig.~\ref{c1c3}) \cite{MAW1,MAW2}.
In the phase chaos regime states with nonzero average phase gradient 
$\nu$ have a dynamics quite different from that at $\nu \sim 0$.
In particular, these states can be either chaotic or regular
depending on the initial conditions and on the parameters 
$c_1,c_3$ and $\nu$. 
In this paper we will focus on MAWs with $\nu \ne 0$ and on the 
dynamical regime associated to them, that is referred to as ``wound-up'' 
phase chaos \cite{miguel}.
It will be shown that MAWs and wound-up phase chaos exist between the 
dashed and the full curve in Fig.~\ref{c1c3}.


In Section \ref{maw} 
the analysis of MAWs introduced in \cite{MAW1,MAW2} 
is extended to arbitrary values of the 
average phase gradients of the field.
The two parameter family of MAW solutions is parametrized by
the spatial period $P$ of the modulation and by 
the average phase gradient 
\begin{equation}
\nu := \frac{1}{P} \int_0^P dx \varphi_x ~.
\end{equation}
For plane wave solutions,  $\nu$ equals the wavenumber $q$.
In analogy, the phase gradient $\varphi_x$ is often called 
``local wavenumber''.
A linear stability analysis will show that MAWs with $\nu \ne 0$ 
can be  stable even in infinitely  large systems.
In contrast,  MAWs with  $\nu=0$ are always unstable in
systems of large length $L \gg P$.
For systems with periodic boundary conditions the average
phase gradient of the whole system can only be changed, if a space-time defect
occurs : $|A(x,t)|$ drops to zero and $\varphi_x$ locally diverges at a
defect.
Persistent phase chaos with conserved $\nu\le\nu_M\ne0$ has been observed in
numerical simulations of the CGLE (\ref{cgle}) \cite{miguel,torcini}. 
The maximum conserved average phase gradient $\nu_M$ decreases as function of
the coefficients $c_1, c_3$ \cite{miguel,torcini} and vanishes at
the apparent transition from phase to defect chaos.
$\nu_M$ was therefore suggested \cite{torcini} as an order parameter for this
transition.
We extended the numerical determination of $\nu_M$ towards smaller $c_3$ and
report the corresponding data in Fig.~\ref{FnuM}.

In Section~\ref{nuM}, a nonlinear analysis of the Eckhaus instability
allows estimates for which parameter values defects occur.
Lower and upper bounds for the limit $\nu_M$ of wound-up phase chaos 
are derived from the existence and stability properties of the MAWs. 
For increasing $\nu$ the degree of chaoticity associated with the 
wound-up phase chaos decreases \cite{torcini}.
For large enough $\nu$ the dynamics can even become 
regular and stable quasi-periodic MAWs appear.
The diamonds in Fig.~\ref{FnuM} indicate this stability limit for
numerical simulations with fixed $c_1=3.5$.
The analysis in Subsection~\ref{nu1instab} will clarify this 
observation.

The large 
number of parameters ($c_1, c_3, \nu, P$) calls for restrictions.
We limit our analysis to fixed $c_1=3.5$ since most previous numerical work 
has been done at this value \cite{torcini,egolf}.
The results will be presented as projections of the 
$P$ direction onto the ($c_3 , \nu$) plane as well as in cuts through
the parameter space spanned by $c_3, \nu$ and $P$.
Additional investigations of the existence domains of MAWs revealed 
qualitatively similar results for fixed $c_1=0.4, 1.2, 2.1$ 
and $5$ and variable $c_3$ 
as well as for fixed $c_3=0.83$ and variable $c_1$. 
Two of these choices were studied by numerical simulations  in \cite{miguel}.
A similarity transformation maps coherent structures onto each other 
along curves $(c_1+c_3)/(1-c_1 c_3)=const$ in coefficient space
\cite{review}.
The parameters $\nu, P, \omega, v$ are transformed accordingly.
One can thereby extend the results presented here to other values of the 
coefficient $c_1$.
Section \ref{nu1exp} discusses possible observations of MAWs in 
experimental systems.
Finally, Section \ref{nu1disc} summarizes the main results.

\section{Existence and stability of  modulated amplitude waves}
\label{maw}

\subsection{Coherent structure approach}

In analogy with the linear analysis of the Eckhaus instability 
of plane waves we make the following ansatz for saturated modulations
\begin{equation}
A(x,t)= a(z) \mbox{e}^{\displaystyle i \tilde{\phi} (z)} 
\mbox{e}^{\displaystyle i (q x - \tilde{\omega} t)}
\label{nu1ansatz0}
\end{equation}
and rewrite it as 
\begin{equation}
A(x,t)= a(z) \mbox{e}^{\displaystyle i \phi (z)} 
\mbox{e}^{\displaystyle i \omega t}~,
\label{nu1ansatz}
\end{equation}
where $a$ and $\phi$ are real-valued functions of $z\!:=\!x - vt$ and 
$\phi (z)=\tilde{\phi} (z)+q z, \omega=q v-\tilde{\omega}$.
Here $a(z)$ and $\phi(z)$ represent coherent structures \cite{saar}. 
Coherent structures have been studied extensively
\cite{torcini,MAW1,MAW2,saar,mvh} and play an important role in various regimes
of the CGLE \cite{janiaud,pumir,miguel,torcini,MAW1,MAW2,saar,mvh,mm}.

Substitution of ansatz~(\ref{nu1ansatz}) into the CGLE~(\ref{cgle}) 
yields the set of three coupled nonlinear ordinary differential equations
(ODEs) 
\begin{eqnarray}
a_z &=& b \nonumber \\ 
b_z &=& \psi^2 a- \gamma^{-1} [(1+c_1 \omega)a+v(b+c_1\psi a)-(1-c_1 c_3)a^3]
\label{ode} \\
\psi_z &=& -2b \psi/a+ \gamma^{-1} [c_1-\omega+v(c_1b/a-\psi)-(c_1+c_3) a^2]
\nonumber
\end{eqnarray}
where $b\!:=\!a_z, \psi\!:=\!\phi_z$ and $\gamma\!:=\!1+c_1^2$
\cite{note1}.
The continuation software {\sc AUTO97} \cite{auto97} is used to compute the
periodic orbits of the ODEs (\ref{ode}) that correspond to spatially periodic
functions $a(z), \phi(z)$. 
In order to choose a unique solution from the
continuous two-parameter family of periodic orbits we set the system size $L$ 
equal to the period $P$ of the periodic
orbit and fix its average phase gradient by $\nu = \frac{1}{L} \int_0^L \psi dz $~.

\begin{figure}
\begin{center}
\epsfxsize=0.9\hsize \mbox{\hspace*{-.06 \hsize} \epsffile{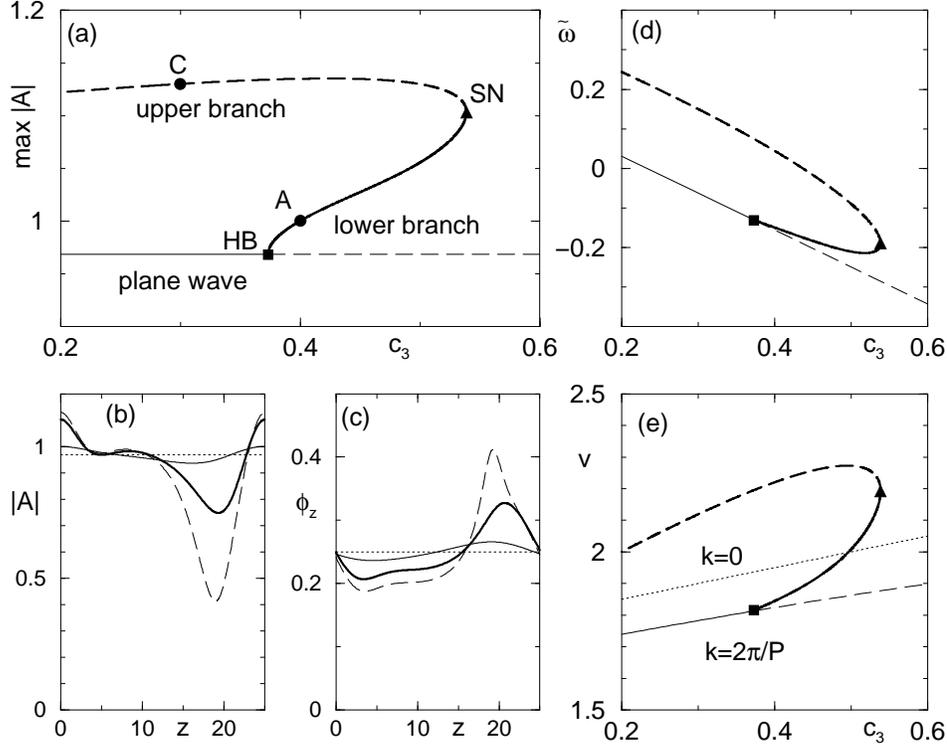} }
\end{center}
 \caption[Bifurcation diagram for $\nu = 0.25, P=25.13, c_1=3.5$]
 {(a) Example of a bifurcation diagram showing the maximum of the modulus 
 for MAWs with $\nu = 0.25, c_1=3.5, P=2\pi/\nu=25.13$. 
 The plane wave that is stable (unstable) against modes of wavelength $P$ is
 represented by the thin full (dashed) line. The stable lower branch (unstable
 upper branch) of MAWs is denoted by the thick full (dashed) curve.
 HB denotes the Hopf bifurcation (square) of the plane wave
 solution whereas SN stands for the saddle-node bifurcation (triangle) that limits
 the existence of MAWs.
 Spatial portraits of (b) the modulus and (c) the phase gradient 
 are shown for a choice of
 solutions. The dotted line represents the plane wave whereas thin full and
 thick full curves give MAWs at locations labelled by A and SN in
 (a).
 The dashed curve denotes the saddle-type upper branch solution at C in (a).
 (d) shows the oscillation frequency $\tilde{\omega}=q v-\omega$ and
 (e) the velocity $v$ versus $c_3$.
 In (e) the dotted line denotes the group velocity
 ($k=0$) and the line below gives the velocity $v_c$ corresponding
 to the mode with finite wavelength $P$ \cite{notevHB}.
 }
 \label{q25L1n1bif}
\end{figure}

The continuation procedure starts from a fixed point $(a,b,\psi) =
(\sqrt{1-q^2},0,q)$ that corresponds to a plane wave solution.
Varying $c_3$, a Hopf (HB) bifurcation (filled square in Fig. 
\ref{q25L1n1bif}) is detected in the ODEs 
where the mode with the smallest possible wave number 
$k_{HB}=2\pi/P $ destabilizes the plane wave. 
Continuing the resulting branch of MAWs the free parameters 
$\omega$ and $v$ are 
adjusted by the continuation algorithm. 
The continuation follows a unique branch of MAWs with $\nu=q$ and $P=L$.
Fig.~\ref{q25L1n1bif} shows examples of resulting bifurcation diagrams.

\begin{figure}
\begin{center}
 \epsfxsize=0.7\hsize \mbox{\hspace*{-.06 \hsize} \epsffile{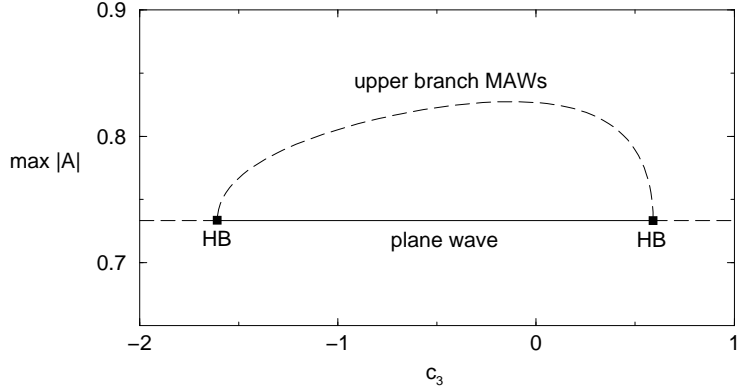} }
\end{center}
 \caption[Bifurcation diagram for subcritical Eckhaus instability]
 {Bifurcation diagram with exclusively upper branch MAWs for 
 $\nu=0.68, P=2\pi/\nu\approx9.24$.
 The solid line indicates stable and the dashed curves unstable
 solutions. The Hopf bifurcations are both subcritical.
 }
 \label{q68L1n1bif}
\end{figure}

\subsection{Existence limits of MAWs}
\label{nu1bif}

Upon increasing of $c_3$ amplitude modulations grow and develop a 
localized depression $|A|_{min}$ where $\phi_x$ has a maximum 
(see Fig.~\ref{q25L1n1bif}b,c).
As for $\nu=0$, these MAWs are called the {\it lower} branch 
in contrast to the coexisting {\it upper} branch MAWs.
The upper branch MAWs are always unstable,
while the MAWs of the lower branch can be stable in
appropriate parameter regions.
Examples of these lower branch MAWs have been obtained by 
numerical simulations earlier 
\cite{janiaud,pumir,miguel,torcini}. 
They have been analyzed in detail in \cite{hager}.
Numerical simulations can neither uncover unstable upper branch MAWs
nor elucidate the limits of existence of MAWs.
The bifurcation analysis presented here reveals that 
upper and lower branch meet and terminate in a saddle-node 
(SN) bifurcation (filled triangle in Fig.~\ref{q25L1n1bif} a,d,e ). 
Due to the SN bifurcation the upper branch MAWs always have at least one 
unstable eigenmode, see also  \cite{mvh,mm}.
The upper branch continues to negative $c_3$ and there connects to
another instability of the plane wave with identical $\nu$ and $ P$.
In the following we will concentrate on the lower branch MAWs.

\begin{figure}
\begin{center}
 \epsfxsize=0.7\hsize \mbox{\hspace*{-.06 \hsize} \epsffile{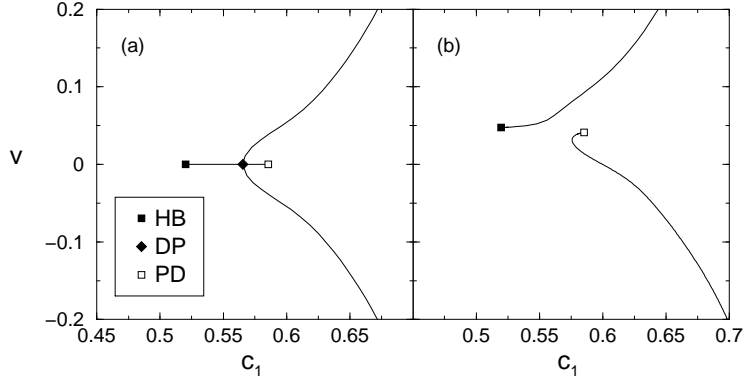} }
\end{center}
 \caption[Unfolding of drift pitchfork bifurcation]
 {Bifurcation diagrams showing the velocity $v$ versus $c_1$. 
 (a) Branches with $v\ne 0$ emerge at the drift pitchfork bifurcation (DP) for
 $c_3=2, P=25, \nu=0$.
 (b) The bifurcation is unfolded for $\nu\ne0$, here $c_3=2, P=25, \nu=0.01$.
 An equivalent pair of branches exists for $\nu\to-\nu$ and $v\to-v$.
 }\label{dpunfold}
\end{figure}

For large $\nu$ and small $P$, the Hopf bifurcation is no longer supercritical
and an unstable branch emerges directly from the plane wave. 
This is in agreement with analytical predictions \cite{janiaud}.
Fig.~\ref{q68L1n1bif} shows an example which also includes the second HB
at negative $c_3$.
For $\nu=0$ the MAWs emerge stationary \cite{MAW1,MAW2} and acquire $v\ne0$ 
above a subsequent drift pitchfork bifurcation \cite{DwightDP}. 
In the present case $\nu\ne0$ the plane wave 
already breaks the reflection symmetry, the initial MAW
has a nonzero velocity and the drift pitchfork (DP) bifurcation 
(filled diamond) is unfolded. 
See Fig.~\ref{dpunfold} for an example at fixed $c_3=2$.
The branch emerging at the HB in Fig.~\ref{dpunfold}b represents the
MAWs as discussed above.
The second branch in Fig.~\ref{dpunfold}b emerges at the period doubling (PD) 
bifurcation (open square) of MAWs with half
the period. 
It always has unstable eigenmodes that drive the dynamics away 
from it to the coexisting MAWs of shorter period. 
Therefore this (upper) branch plays no
essential role and is not treated further.

\subsubsection{Infinite system size}

\begin{figure}
\begin{center}
 \epsfxsize=0.7\hsize \mbox{\hspace*{-.06 \hsize} \epsffile{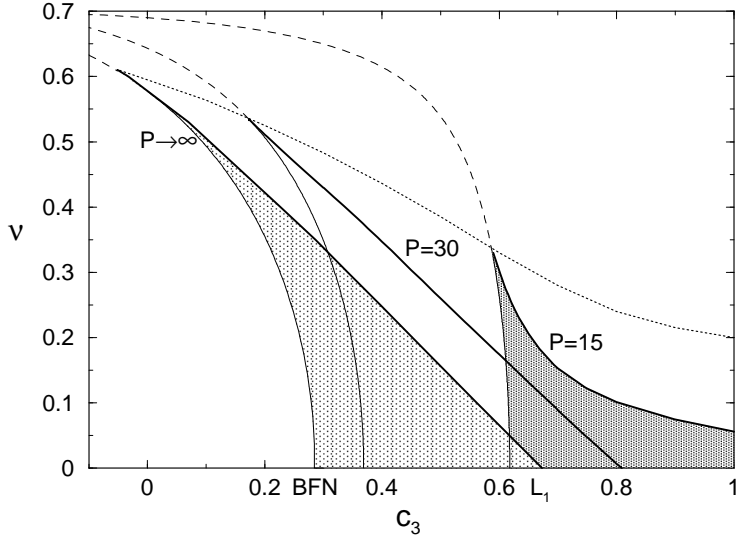} }
\end{center}
 \caption[Existence domains of MAWs in $L\to\infty$]
 {Existence domains of MAWs with period $P$ projected onto the 
 ($c_3, \nu$) parameter plane.
 Thin curves denote the Eckhaus instability and HB which occur supercritical 
 (full curve) or subcritical (dashed) depending on $\nu$ and $P$.
 The thick curves give the SN for selected periods $P$.
 Three examples of existence domains for
 $P=15$ (right, dark shaded domain), 
 $P=30$ (middle, empty), 
 $P\to\infty$ (left, light shaded) are shown.
 The superposition of all existence domains is bounded by the dotted curve.
 }
 \label{cqexist}
\end{figure}

We have analyzed the existence of lower branch MAWs in the entire parameter space
($c_3, \nu, P$) at fixed $c_1=3.5$. The system size is assumed infinitely
large in order to allow for arbitrary periods $P$ of MAWs.
Fig.~\ref{cqexist} shows examples of existence
domains for $P=15, P=30$ and $P\to\infty$. 
We find that both HB and SN shift to
larger $c_3$ as the period $P$ is decreased. 
The same behavior has already been observed 
in the special  case $\nu=0$ \cite{MAW1,MAW2}. 
The dotted curve in Fig.~\ref{cqexist} indicates the
``envelope'' of all SN bifurcations for MAWs of any
period and therefore is the upper boundary for the
existence domain of the MAWs.

\subsubsection{Medium system size}

Experimental setups and numerical simulations are restricted to finite system
size $L$.
Often periodic boundary conditions (corresponding to an annular geometry) 
are used in order to study bulk effects of extended systems and to 
minimize boundary effects.
The periodic boundary conditions also restrict possible modes of
perturbations.
As described by Eq.~(\ref{sigmak}) the instability
threshold of plane wave solutions depends on the wavenumber $k$ of the
perturbation. Since in the studied range of coefficients the Eckhaus instability
is a long-wavelength instability the plane waves will be stabilized 
in small systems. 
The instability threshold is shifted to larger values
of the coefficients $c_1, c_3$ and can be computed from
Eq.~(\ref{sigmak}) setting $k=2\pi/L$.

\begin{figure}
\begin{center}
 \epsfxsize=0.9\hsize \mbox{\hspace*{-.06 \hsize} \epsffile{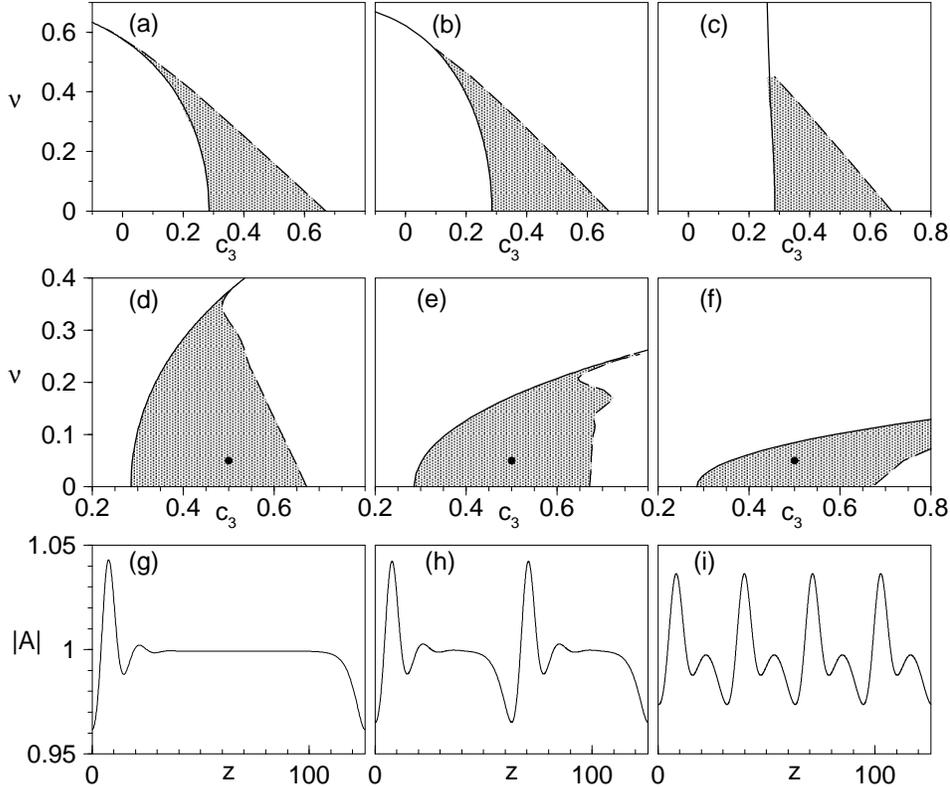} }
\end{center}
 \caption[Existence domains of MAWs]{
 Existence domains of lower branch MAWs are denoted by shaded areas for
 (a) $n=20$, (b) $n=4$, (c) $n=2$, (d) $n=1$, (e) $n=1/2$ and (f) $n=1/4$. 
 They are limited by HB (solid curve) at small $c_3$ and by the SN 
 (dashed curve) at large $c_3$.
 Spatial profiles of coexisting MAWs at $\nu=0.05, c_3=0.5$ are shown for 
 (g) $n=1$, (h) $n=1/2$ and (i) $n=1/4$, corresponding to dots in (d-f).
 }
 \label{mawn}
\end{figure}

Clearly the selection of perturbations by periodic boundary conditions also
restricts possible MAWs. 
Their average phase gradient $\nu$ and the
period $P$ have to be consistent with the system size and this
renders the two-parameter family of MAWs discrete. 
It is thus convenient to parametrize MAWs by the average phase gradient
$\nu$ and the ratio $n$ of wavelength
\begin{equation}
n := \frac{P}{2\pi/\nu} ~.
\label{n}
\end{equation}
The ratio $n$ takes values of integer fractions where the nominator counts the
number of underlying wave length and the denominator the number of 
humps of the modulation. Hence this quantity is easily accessible in experiments.
The existence domains of MAWs with respective $n$ are 
presented in the ($c_3, \nu$) parameter plane in 
Fig.~\ref{mawn}.

\subsubsection{Small system size}

\begin{figure}
\begin{center}
 \epsfxsize=0.9\hsize \mbox{\hspace*{-.06 \hsize} \epsffile{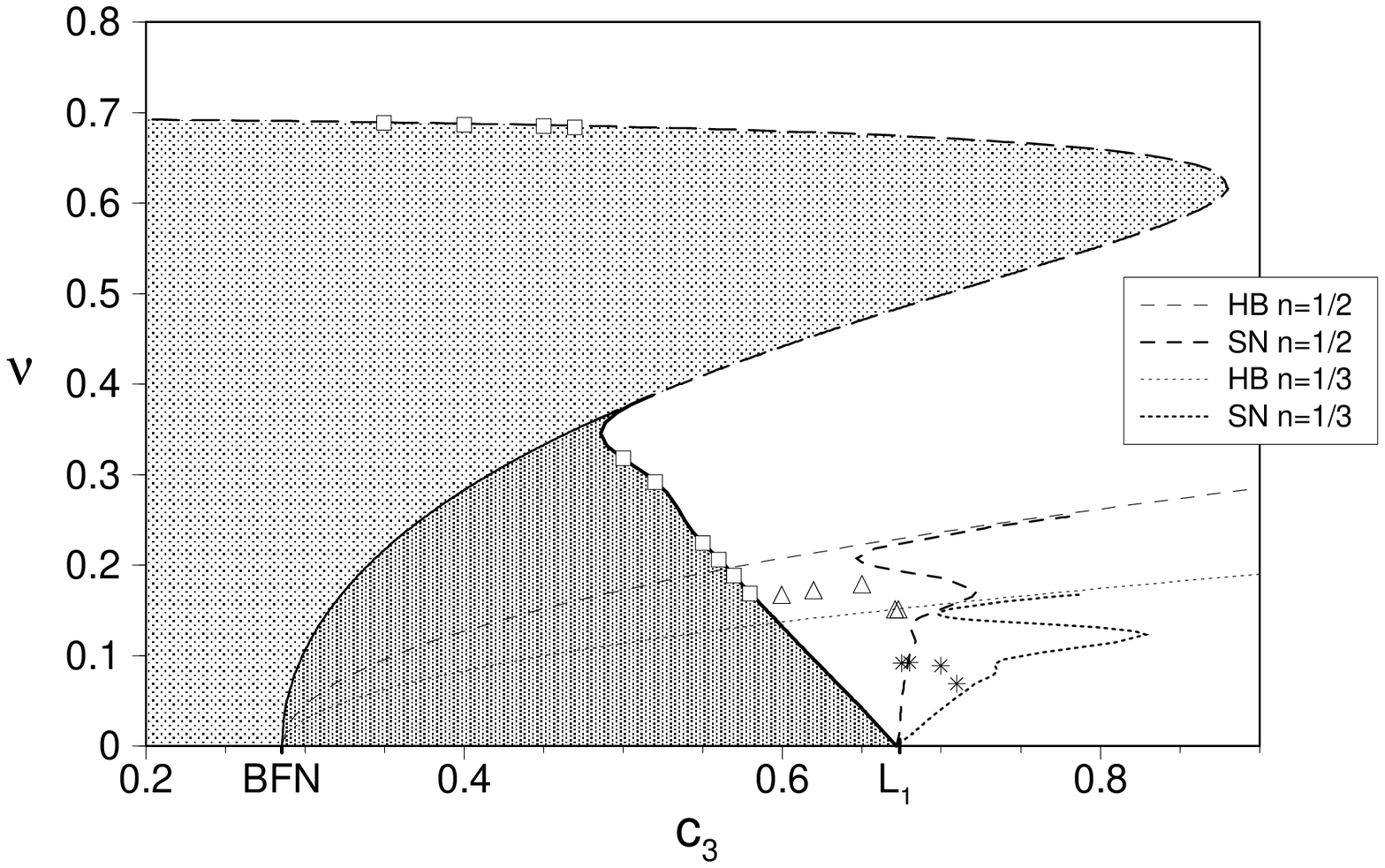} }
\end{center}
 \caption[Existence domains of MAWs in $L=2\pi/\nu$]
 {For short system size $L=2\pi/\nu$ plane waves are stable for parameter choices
 inside the light shaded area.
 MAWs with a single hump ($P=L$) exist inside the dark shaded area bounded by 
 the supercritical HB to the left and the SN to the right.
 Thin curves give the limits of MAWs with two humps $P=L/2$ (dashed)
 and three humps $P=L/3$ (dotted). See the legend for the different cases.
 Symbols denote maximal $\nu=\nu_U$ that did not create defects but resulted in stable
 asymptotic states in simulations of the short system.
 Plane waves and single MAWs (squares), multi-hump MAWs with 2 humps 
 (triangles) and 3 humps (stars) have been observed at $\nu=\nu_U$.
 Data $\nu_U$ are taken from \cite{torcini}.
 } 
 \label{L1}
\end{figure}

Here we focus on the extreme case. 
The shortest possible system 
with periodic boundary conditions only contains one wavelength of the 
plane wave, consequently its length $L$ is given by  $L=2\pi/\nu$. 
In \cite{torcini} the quantity $\nu_U$ was determined 
in analogy to $\nu_M$ for large systems. 
$\nu_U$ denotes the largest $\nu$ for which none of the random initial
conditions (different realizations of noise added to a plane wave) 
produced a defect. 
In the following these data (symbols in Fig.~\ref{L1}) are compared to the 
existence domains of MAWs.

Within the light shaded area in Fig.~\ref{L1} plane wave solutions with
wavenumber $\nu$ are stable in the short system. 
The stability area extends over the phase chaos and into the defect chaos 
region. 
We stress this result because from the 
experimental observation of stable plane waves one cannot necessarily 
infer that the dynamics of the system may be reproduced by the CGLE
with coefficients $c_1, c_3$ in the Benjamin-Feir stable region.
The dashed curve denotes a subcritical instability.
Only unstable upper branch MAWs exist to the left of this curve. 
For smaller $\nu$ the instability again
turns supercritical and stable lower branch MAWs exist
inside the dark shaded region. 
The thick full curve gives the SN bifurcation 
for MAWs with $P=L$. 
The thin curves show the respective limits of MAWs with shorter
period. 
Defects are expected beyond the SN \cite{MAW1,MAW2} and the subcritical
instability which well reproduces the data from numerical simulations
\cite{torcini} except at small $\nu$.
Simulations with 
$\nu \le \nu_U$ resulted in modulations with a single hump
(squares) or with two (triangles) or three (stars) humps of different size
\cite{torcini}.
The latter two are observed above the SN of MAWs with $P=L$. 
Here the initial conditions select
MAWs with shorter period which only coexist at small $\nu$. 
The SN with $P=L$ nevertheless gives a lower bound for the formation
of defects.  

\subsection{Instabilities of MAWs}
\label{nu1instab}

\begin{figure}
\begin{center}
 \epsfxsize=0.7\hsize \mbox{\hspace*{-.06 \hsize} \epsffile{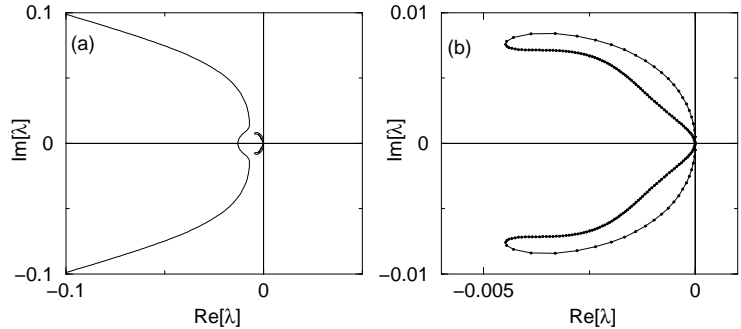} }
\end{center}
 \caption[Stable MAW in $L\to\infty$]
 {
 (a) Spectrum of eigenvalues $\lambda$ of a stable lower branch MAW. 
 Parameters are $c_3=0.4, \nu=0.184, P=2\pi/\nu$.
 (b) Blow-up of the leading part of the spectrum.
 The dots correspond to system size $L=100*P=3415$ and have been calculated
 using the Bloch method \cite{MAW2,bloch}.
 } 
 \label{L_infty}
\end{figure}

\begin{figure}
\begin{center}
 \epsfxsize=0.9\hsize \mbox{\hspace*{-.06 \hsize} \epsffile{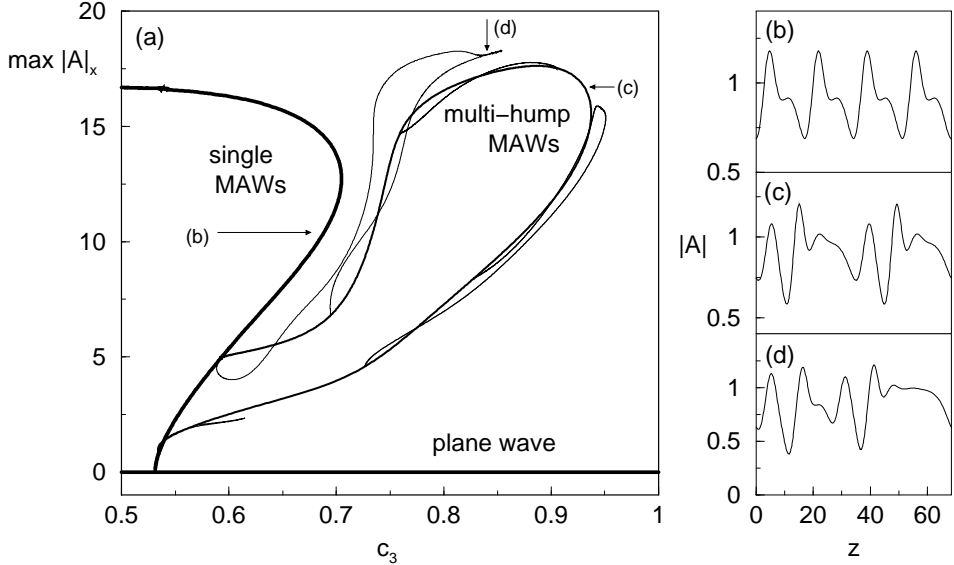} }
\end{center}
 \caption[Instability : multi-hump MAWs]
 {(a) Bifurcation diagram for $\nu=0.184, P=\pi/\nu$ and $L=4*P=68.3$ hence 4
 pulse-like modulations interact. 
 The maximum of the amplitude gradient is plotted since the
 interaction causes pulse shifts and the amplitude of single humps changes
 little. Thicker lines correspond to smaller overall period of the
 modulation. 
 Typical solutions are shown in (b)-(d) as indicated by arrows in (a).
 } 
 \label{L4pd}
\end{figure}

\begin{figure}
\begin{center}
 \epsfxsize=0.7\hsize \mbox{\hspace*{-.06 \hsize} \epsffile{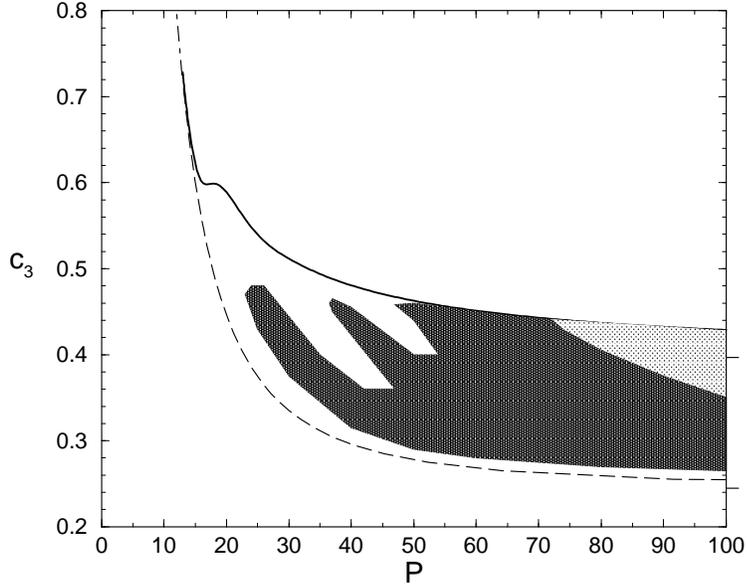} }
\end{center}
 \caption[Stability domains of MAWs for $\nu=0.25$]
 {Stability domain (dark area) of single MAWs for $\nu=0.25, L\to\infty$.
 MAWs exist between supercritical HB (dashed curve) and SN (full curve). 
 The tick marks at the right frame give the asymptotic values for $P\to\infty$. 
 The dot-dashed curve denotes the subcritical HB.
 MAWs are unstable to splitting within the light shaded domain at large $P$. 
 Within the white domain at small $P$ single MAWs are unstable to 
 interaction.
 } 
 \label{nu1q25}
\end{figure}

\begin{figure}
\begin{center}
 \epsfxsize=0.7\hsize \mbox{\hspace*{-.06 \hsize} \epsffile{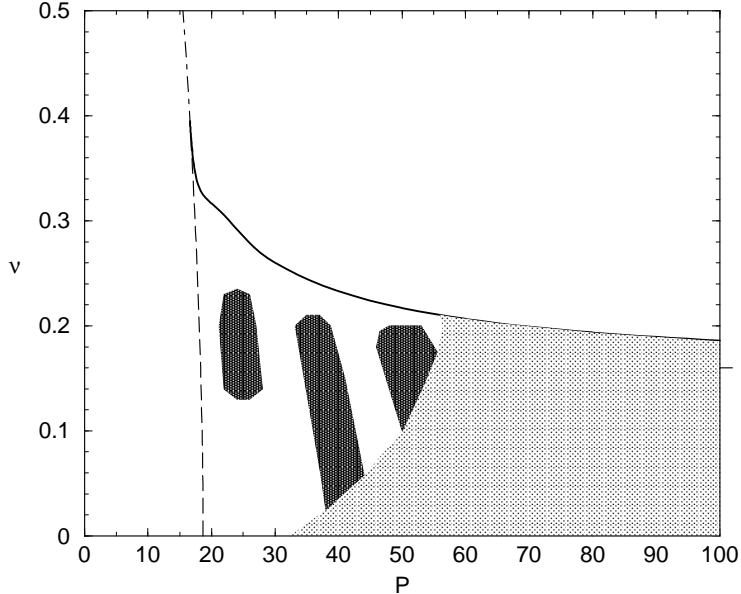} }
\end{center}
 \caption[Stability domains of MAWs for $c_3=0.5$]
 {Stability domains (dark areas) of single MAWs for $c_3=0.5, L\to\infty$.
 Curves and instability domains (splitting=light shaded, interaction=white) have
 the same meaning as in Fig.~\ref{nu1q25}.
 } 
 \label{nu1c5}
\end{figure}

In contrast to the case of MAWs with zero average phase gradients 
some MAWs with non-zero $\nu$ are stable even in very large systems. 
A linear stability analysis of MAWs as in \cite{MAW2} yields the spectrum of
eigenvalues as shown in Fig.~\ref{L_infty} for a typical example.
From Fig.~\ref{L_infty} we conclude that for this example the entire spectrum in 
the infinite system will be confined to the left half-plane.
Thus MAWs should be found in experiments, that can be well described 
by the CGLE for appropriate control parameters. 
In this section we present a detailed study of the stability
properties of MAWs. 
MAWs with a single hump per period $P$ will be called ``single MAWs''.
Their existence domains were studied in the previous section.
However, the effective interaction between adjacent periods of a single MAW 
can be repulsive or attractive (see Figs. \ref{nu1q25} and \ref{nu1c5}). 
Period doubling (PD) bifurcations (open squares) occur at the transitions
from repulsive to attractive interaction \cite{michal}.
There, new branches of MAWs with longer period but many humps per period
emerge from the primary branch of single MAWs.
We will call these solutions ``multi-hump MAWs''. 
In their profile some humps gain more space and others are compressed in an
alternating fashion.
The new branches extend to larger $c_3$ than the corresponding single MAWs.
Fig.~\ref{L4pd} shows how these branches arrange in a system
with 4 interacting humps ($L=4*P$).
As long as the PD bifurcations are supercritical,
the multi-hump MAWs are stable. 
They represent the
saturated solution for attractive interaction between subsequent modulations.
For large systems a whole sequence of period doubling bifurcations will lead 
to multi-hump MAWs with an overall period equal to the system size. 
Hence they appear as an 
erratic spatial sequence of humps and depressions. 
This spatial sequence propagates in a coherent fashion.
We named these patterns multi-hump MAWs to emphasize the 
connection among the coherent structures.

Examples of these stable aperiodic patterns 
were already observed in numerical simulations. 
R. Montagne {\it et al.} \cite{miguel} denote this behavior as
``frozen phase turbulence'' while
A. Torcini {\it et al.} \cite{torcini} use the term ``solutions of 
type $\beta$''. 

The observed coexistence of a large number of stable multi-hump MAWs results 
in a strong dependence of the final state on the initial 
conditions of the numerical simulation. 
Although each regular final configuration must be consistent with
a particular single or multi-hump MAW it is difficult to predict how 
the selected final patterns depend on the initial conditions. 

MAWs with large period $P$ undergo a ``splitting'' instability as in the limit
case $\nu=0$ \cite{MAW2}. Roughly, the spatial profiles of these MAWs consist of
a localized hump and a plane wave part. Since the extended plane wave is
linearly unstable the splitting instability is reminiscent of the Eckhaus
instability.
It creates more humps on the plateau of the unstable
MAW and reduces the period $P$ of MAWs on average.

Figs.~\ref{nu1q25} and \ref{nu1c5} represent cuts through the parameter space at
fixed $\nu=0.25$ and $c_3=0.5$, respectively. They show the typical arrangement
of stable and unstable parameter regions for single MAWs. 
Other examined cuts for
$c_3=0.1, 0.2, 0.3, 0.4, 0.6$ and $0.7$ qualitatively show the same order.

The cut through parameter space $c_3, \nu, P$ at $\nu=0.25$ is shown in
Fig.~\ref{nu1q25}. The HB (dashed curve in the figure) approaches the 
Eckhaus instability for 
$P\to\infty$ as the lower bound of the existence domain. 
From above the domain is limited by the SN (solid curve). 
For small $P$ (large $c_3$) the HB is subcritical and 
only unstable upper branch MAWs exist to the left.
In the infinite system MAWs are found to be linearly stable for
a broad range of parameters (dark shaded area). 
At low $P$ the interaction instability occurs (white area)
whereas at large $P$ the long plateau of the MAW is unstable to splitting (light
shaded area).
For $c_3<0.45$ most random initial conditions will evolve to stable MAWs. 

A cut perpendicular to the previous one is shown in Fig.~\ref{nu1c5}.
Curves and shadings have the same meaning as discussed above. 
Starting from random initial conditions at $\nu>0.1$ a transient
may again lead to a stable MAW with local 
periods $P$ inside the stable windows. 
At lower $\nu<0.1$ the probability
of approaching a stable configuration decreases 
since only a third of the previous stable $P$ intervals remains. 
Below $\nu=0.02$ no stable state can be prepared at all.
Instead one observes wound-up phase chaos 
with an associated maximal Lyapunov exponent that
increases for decreasing $\nu$.
As in the limit case $\nu=0$ (phase chaos), the dynamics is driven by the
attractive interaction and annihilation of 
localized modulations in competition with the splitting instability 
that produces new peaks in the modulations.
In particular for decreasing $\nu$, 
the splitting instability extends to shorter periods $P$
and significantly overlaps with the interaction
instability.
With the above arguments many results obtained by numerical simulations 
of the CGLE can be well interpreted. In particular it has been observed 
in \cite{torcini} for the same choice of parameters ($c_1=3.5$ and $c_3=0.5$) 
that the maximal Lyapunov exponent (averaged over many different initial
conditions) is positive for $\nu=0$ and decreases monotonously 
towards zero for increasing $\nu$. Above $\nu=0.09$ no chaotic solutions
have been observed.

\section{Defect formation in wound-up phase chaos}
\label{nuM}

In this section the formation of defects and the resulting change of the 
average phase gradient are studied. 
For $\nu\ne0$, the scenario of defect formation past the
saddle-node bifurcation of the relevant MAW is analogous to the
previously studied case $\nu = 0$  \cite{MAW1,MAW2}, see Subsection
\ref{nu1sn}.
In particular, the dependence of the final selected average phase gradient
$\nu_f$ on the initial value $\nu_i$ found in numerical simulations 
\cite{miguel} can be interpreted.
Subsection \ref{nuMok} is then devoted to the limit $\nu_M$ of wound-up phase
chaos. 
For a certain range of parameters the limit $\nu_M$ is reproduced by
means of the stability properties of MAWs at the saddle-node
bifurcation. 
These arguments work well for defect creation with $\nu > 0.1$. 
At smaller values of $\nu$ various instabilities (splitting and
interaction) of MAWs compete and a general statement is more
difficult, compare also \cite{MAW2}.

\subsection{Beyond the saddle-node bifurcation}
\label{nu1sn}

\begin{figure}
\begin{center}
 \epsfxsize=0.9\hsize \mbox{\hspace*{-.06 \hsize} \epsffile{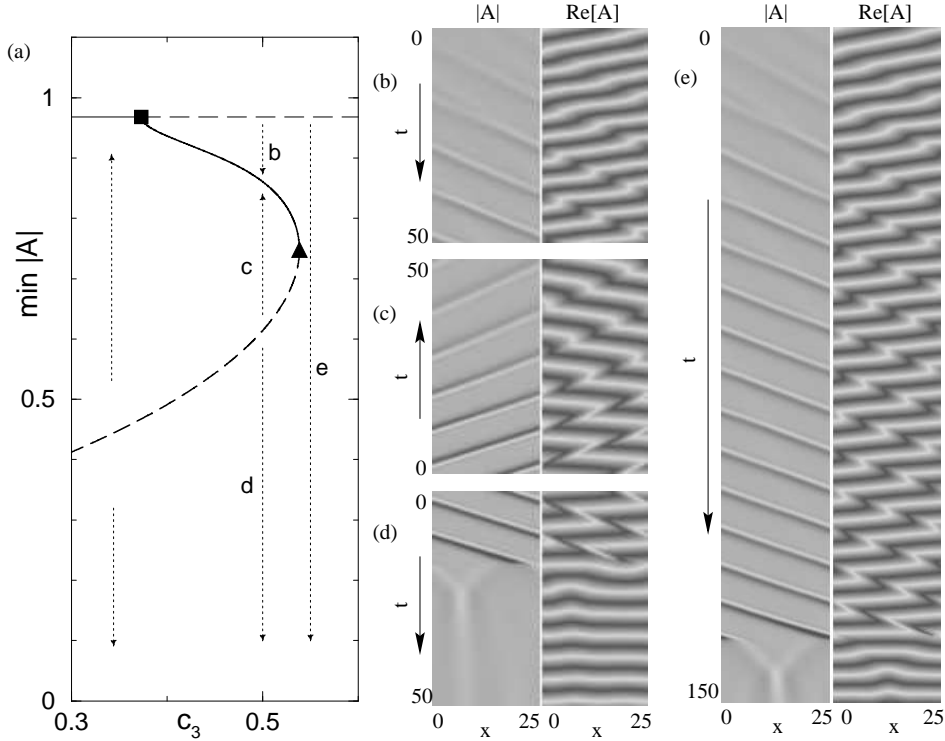} }
\end{center}
 \caption[Defect formation : dynamics along SN manifold]
 {(a) Bifurcation diagram as in Fig.~\ref{q25L1n1bif} showing the minimum modulus of the
 MAWs.
 (b)-(e) Numerical simulations illustrate the dynamics near the SN 
 corresponding to the arrows in (a).
 (b) Plane wave perturbed at one point and
 (c) unstable saddle-type MAW plus noise converge to the stable MAW.
 (d) Unstable saddle-type MAW plus a different realization of noise 
 evolves to a defect that changes $\nu$ to 0.
 (e) As (b) but beyond the SN which makes defect formation
 possible for arbitrarily small perturbations of the plane wave.
 Note the long living transient of a non-coherent modulation.
 (b)-(d) are at $c_3=0.5$ below the SN and
 (e) belongs to $c_3=0.55$ above the SN for $\nu=0.25,
 P=L=2\pi/\nu=25.13$.
 }
 \label{nu1snsim}
\end{figure}

The role of the SN bifurcation for the dynamics has been studied in
\cite{MAW1,MAW2} for the limit case $\nu=0$. For $\nu\ne0$ 
we find similar behavior. 
Fig.~\ref{nu1snsim} gives examples
for $\nu=0.25, P=2\pi/\nu$.
Perturbations of a plane wave lead to defects only above the SN, 
whereas below the SN such perturbations have to be very large to overcome the 
saddle-type upper branch MAW. 

There are no SNs for parameters below the SN corresponding to $P\to\infty$.
Thus, starting from random initial conditions defects may only form at 
parameters above the SN of $P\to\infty$. 
The SN of $P\to\infty$ represents a lower bound for defect formation.

For large systems the formation of defects depends on the local period of
initial perturbations in a similar way as for $\nu=0$ \cite{MAW1,MAW2}. 
The peak to peak distances of $\varphi_x(x,t)$ are used to determine local
periods $p$.
In that context, the following explanation of defect 
formation has been proposed. Defects are observed in the
phase turbulent regime whenever local structures, similar to MAWs,
with spatial periods $p$ larger then $P_{SN}$ occur in the system.
Where $P_{SN}$ denotes the period of the MAWs at the SN
(that coincides with the maximal MAW-period)
for the considered choice of parameters.
For larger values of $\nu$ or $c_3$ the SN occurs for smaller $P_{SN}$ 
as shown in Figs.~\ref{nu1q25} and \ref{nu1c5}. 
Therefore, at larger $\nu, c_3$
local periods $p$ beyond the SN and
subsequent defect formation are more probable.
In contrast to the case $\nu=0$, there is only a short transient of 
phase chaos in the simulations with nonzero initial $\nu_i > \nu_M$.
The distribution of local periods $p$ of the perturbations is given by the
realization of the noise in the initial condition.
For local periods slightly above the relevant SN 
(as in Fig.~\ref{nu1snsim}e), 
the perturbation first increases to a modulation similar to MAWs and 
appears almost
saturated for some transient time until finally a defect appears. 
This transient of defect formation becomes shorter as the distance to
the SN grows.
If initial conditions are prepared with $\nu_i \gg \nu_M$, then some local
periods will be far beyond the corresponding SNs. 
The transients of defect formation are shorter in this case.

These two observations suggest an interpretation of the curve $\nu(t)$
representing the temporal evolution of $\nu$ during transients 
with $\nu_i > \nu_M$ \cite{miguel}.
The larger the initial $\nu_i$ is chosen the smaller is the final value $\nu_f$ .
The time scales of competing processes have to be considered.
Local defect formation will 
not instantaneously effect distant spatial locations 
along the system.
Instead the local change of the
average phase gradient $\nu$ via a defect will take a transient time to relax 
over the entire system.
For $\nu_i$ much larger than $\nu_M$, defect formation happens on a short 
time scale and
independently leads to defects at many different spatial locations
before the relaxation of the decreased average phase gradient $\nu_f
\ll \nu_M$ can stop defect formation.
For $\nu_i$ just above $\nu_M$ defects form slowly and the reduced $\nu_f$ can relax the 
phase gradient at distant locations before other defects occur.

Let us now verify if the mechanisms proposed to explain
defect formation in the phase chaos regime for solutions
with $\nu \approx 0$ still hold for the wound-up phase chaos regime.
For $\nu \approx 0$, defects form if and only if the period $p$
of local structures is larger than $P_{SN}$, where $P_{SN}$ is the 
period for which a SN occurs at the chosen coefficients $c_1$ and $c_3$
\cite{MAW1,MAW2}.
We have considered two cuts in the parameter space at
$c_3=0.5$ and $c_3=0.65$ and investigated the
distributions of periods $p$ for solutions with average phase
gradient $\nu \sim \nu_M$. In particular, 50 realizations
of a system of length $L=512$ initialized as a plane wave
with wavenumber $\nu$ plus noise in the amplitude and in the
phase have been considered. Then these different initial
conditions have been followed for an integration time
$t=500,000-1,000,000$. The last part of the run has
been examined in order to extract the length of the coherent
structure with the maximal period $p_{max}$
occurring during the evolution.

From these simulations we obtain the following: if defects
occur then $p_{max} > P_{SN}$
in all observed cases. However, it is not true that
a defect is formed any time we observe $p_{max} > P_{SN}$.
If we let the system relax for a long time
($t=500,000$) and measure $p_{max}$,
then the number of initial conditions
leading to a $p_{max} > P_{SN}$ without defect formation
is noticeably reduced. For $c_3=0.65$  the
maximal conserved phase gradient is $\nu_M \sim 0.1$ .
In the simulations we do not observe a defect for
$\nu=0.086$ and $\nu=0.098$ but in the first
case only 2 \% of all runs show
$p_{max} > P_{SN}$, while in the latter case this percentage
increases to 8 \%. 
Increasing $\nu$ the maximal period $p_{max}$ increases and
more and more situations with $p_{max} > P_{SN}$ are found
upon approach to $\nu_M$.

The differences to the $\nu=0$ case may be explained by the
coexistence of chaotic and stable not-chaotic attractors.
Depending on the initial conditions, the solution
of the CGLE can evolve towards one or the other.
Therefore the system may exhibit local structures similar to
multi-hump MAWs that possess SN bifurcations at parameter values 
larger than those of single MAWs (compare Fig.~\ref{L4pd}).
In that case, some periods $p$ may even exceed $P_{SN}$.

\subsection{Limit of wound-up phase chaos}
\label{nuMok}

For random initial conditions with $\nu$ in the narrow range between the SN 
of $P\to\infty$ and the existence limit of MAWs (see Fig.~\ref{cqexist}),
it depends on the specific realization of the noise
whether a defect can form or a stable MAW results.
No defects form below the line $\nu_M (c_3)$.
In order to understand this observed limit $\nu_M (c_3)$ of wound-up phase chaos, 
it is sufficient to consider the SNs of single MAWs.

\begin{figure}
\begin{center}
 \epsfxsize=0.9\hsize \mbox{\hspace*{-.06 \hsize} \epsffile{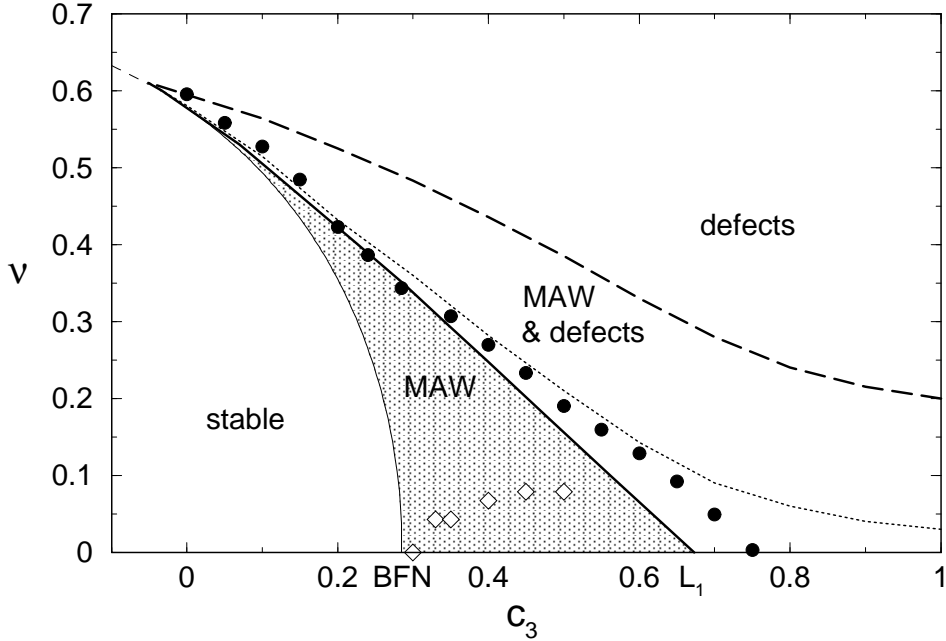} }
\end{center}
 \caption[Theoretical bounds for $\nu_M (c_3)$]
 {Theoretical bounds for $\nu_M (c_3)$ (filled circles) : 
 SN of $P\to\infty$ as lower bound (thick solid curve)
 and the presence of splitting at SN as upper bound (dotted curve).
 Other curves as in Fig.~\ref{FnuM}. 
 See text for details.
 } 
 \label{cqnuM}
\end{figure}

Although initial conditions with large $P$ beyond a SN could lead to defects,
this is often prevented by the action of the splitting instability. 
Then the period is decreased before a defect can form.
Following the SN curve in Figs.~\ref{nu1q25} and \ref{nu1c5} one encounters a
transition from SNs with a splitting instability at large $P$ to SNs without
this instability at short $P$.
Defect formation in wound-up phase chaos with $\nu > 0.1$ occurs for parameters
where the splitting instability is {\it not} present near the SN, {\it
i.e.} above the dotted line in Fig.~\ref{cqnuM}.

Fig.~\ref{cqnuM} summarizes the bounds found so far  for the limit
of wound-up phase chaos.
The domain of stable plane waves at low $c_3$ is limited by the Eckhaus
instability (thin curve). 
Within the shaded area only supercritical HBs of different period $P$ occur 
but no SNs.
This area is limited by the lowest SN curve of $P\to\infty$ (thick
solid curve). 
No defects do form from random initial conditions within the shaded
area.
The dashed curve denotes the upper limit of the existence domain 
of MAWs with any period.
Saddle-node bifurcations exist in the window between this dashed curve and 
the thick solid curve for the SN with $P\to\infty$.
Defects will always form for any choice of initial conditions at parameters 
above the dashed curve.
The dotted curve gives the transition from active (below) to inactive (above)
splitting modes at the SN. 
This transition is computed by linear stability analysis along cuts like
Figs.~\ref{nu1q25} and \ref{nu1c5}.
Splitting inhibits defect formation below this dotted curve.
Filled circles correspond to $\nu_M$ obtained from numerical simulations
as in \cite{torcini}.
Diamonds refer to the transition from chaotic (below) to non-chaotic (above) 
asymptotic states.

The SN for $P\to\infty$ (thick solid) is a lower bound for defect
formation which also holds in the limit $\nu=0$. 
The point L$_1$ marks the transition from phase to defect chaos
studied earlier \cite{MAW1,MAW2}.

As long as the dynamics is regular ($\nu>0.1$), the upper bound for the limit
(onset of splitting at SN, dotted curve) of wound-up phase chaos
reproduces well the numerical observations. 
For chaotic states with $\nu<0.1$ defect formation eventually becomes possible 
even despite the presence of the splitting instability. 
This coincides with the increasing weight of the instability of single MAWs 
to attractive interaction of subsequent amplitude peaks. 

\section{Experimental observations}
\label{nu1exp}

A variety of experimental observations in
quasi one-dimensional geometries can be well interpreted by MAWs.
These systems shall also serve for testing further properties of MAWs
\cite{mvhexp}.
 
({\it i}) 
For the oscillatory instability of Rayleigh-B\'enard convection patterns
in an annular cell, B. Janiaud {\it et al.} report long living transients of 
modulated waves that eventually cause defects \cite{janiaud}. 
The underlying Eckhaus instability was found to be subcritical. 
Then, we expect the modulation to grow roughly exponentially
whereas the long transient of an almost saturated modulation is similar to
the dynamics near the saddle-node bifurcation as shown in
Fig.~\ref{nu1snsim}e.
In the latter case the Eckhaus instability is supercritical and
stable MAWs may exist for nearby values of the experimental
parameters, see also the discussion in \cite{janiaud2}.
In all cases a single pulse-shaped modulation with the period equal to the cell 
length was present in the system and led to a defect. 
The excitation of several modulations per cell
and thereby smaller period can stabilize the modulated pattern and provide more
examples of the dynamics near the saddle-node bifurcation.

({\it ii}) 
Hydrothermal waves have been studied in ring-shaped cells \cite{h2} 
as well as in linear cells \cite{h1,garnier,CIAI4}.
N. Mukolobwiez {\it et al.} report a supercritical Eckhaus instability,
stable MAWs with the period equal to the cell length and defect formation 
after a parameter change \cite{h2}. 
N. Garnier {\it et al.} observe modulated waves with both the wavenumber and the
period of modulations being selected by one of the longitudinal
boundaries \cite{garnier}.
Thermal or mechanical forcing at the boundary may yield more insight into the
multistability of the two-parameter family of MAWs.

({\it iii})
In rotating Rayleigh-B\'enard convection Y. Liu {\it et al.} observe the
subcritical Eckhaus instability of a traveling wave sidewall mode
\cite{sidewall}. 
The authors suggest higher order corrections to the CGLE in
order to explain the observed discrepancy between the linear group velocity and 
the observed velocity of finite wavelength perturbations.
However, this difference already follows from the linear analysis 
of the Eckhaus instability \cite{notevHB} and can be used as a test
for the assumed coefficients of the CGLE.

({\it iv})
For the Taylor-Dean system I. Mutabazi {\it et al.} report a stable MAW that
they called ``triplet state'' because of the length scale ratio of modulation and
underlying wave \cite{dean}. 
They also observe the formation of defects. 
Clearly the triplet state is just one realization of the two-parameter family 
of MAWs.

({\it v})
Finally we mention the heated wire convection \cite{wire}, internal waves
\cite{velarde} and the oscillatory variant of the Belousov-Zhabotinsky reaction
\cite{chem} where further investigations of the observed Eckhaus
instabilities may also reveal MAWs.

\section{Discussion}
\label{nu1disc}

The bifurcation analysis of modulated amplitude waves (MAWs) 
has been extended to nonzero average phase gradient ($\nu\ne0$).
Small amplitude MAWs (``lower branch''-MAWs) of specific spatial
period  $P$ exist between a supercritical 
Hopf bifurcation (HB) and a saddle-node (SN) bifurcation. 
The HB asymptotically reaches the Eckhaus instability from above as $P$ goes to
infinity.
MAWs are a direct consequence of the Eckhaus instability
of plane waves; they are obtained by  a computer-assisted nonlinear analysis
of this instability.
We encounter SNs with decreasing values of $P$ as $c_1, c_3$ and $\nu$ are
increased.
These SNs govern the formation of defects from random initial 
conditions as well as many aspects of the evolution of wound-up phase chaos.
The SNs bound the existence region of MAWs in the
Eckhaus unstable regime.

A linear stability analysis of MAWs revealed that
they can be linearly stable even in systems of infinite size.
These domains are limited by the interaction instability at low and the
splitting instability at high values of the spatial period $P$ of the MAW.
The competition of the two instabilities drives wound-up phase chaos and
determines the degree of chaoticity of the dynamics.
%
%
For fixed coefficients $c_1$ and $c_3$,
the SN associated to $P\to\infty$ occurs at the lowest value of $\nu$.
This SN establishes a ``lower'' bound for defect formation and 
thereby for the limit $\nu_M$ of wound-up phase chaos.
The splitting instability can inhibit defect formation if the SN occurs at 
large $P$ hence defects are created more frequently for parameters above a
second curve where the splitting instability vanishes at the SN.
Thereby an ``upper'' bound for the limit $\nu_M$ of wound-up phase chaos 
is obtained.
Earlier numerical observations on $\nu_M (c_3)$ are well reproduced for
$\nu > 0.1$, respectively small $c_3$. 
For $\nu < 0.1$, the description of phase chaos
relies on considerations similar to those already discussed in the 
limit case $\nu=0$ \cite{MAW1,MAW2}.
Finally, several experimental observations were interpreted in terms of MAWs.

We acknowledge fruitful discussions with M. van Hecke, N. Garnier, F. Daviaud,
W. van de Water and L. Kramer.
AT was partially supported by the MURST-COFIN00 
grant ``Chaos and localization in classical and quantum mechanics''.



\begin{thebibliography}{00}
  
\bibitem{books} Y. Kuramoto, {\em Chemical Oscillations, Waves
and Turbulence} (Springer, 1984, Berlin);
P. Manneville, {\em Dissipative structures and Weak Turbulence}
(Academic Press, 1990, San Diego);
T. Bohr, M. H. Jensen, G. Paladin and A. Vulpiani, {\em Dynamical
systems approach to turbulence} (Cambridge Univ. Press) (1998).

\bibitem{CH} M. C. Cross and P. C. Hohenberg, Rev. Mod. Phys. {\bf
65}, 851 (1993).

\bibitem{eck} W. Eckhaus, {\it Studies in Nonlinear Stability Theory}
(Springer Verlag, New York, 1965).

\bibitem{janiaud} B. Janiaud, A. Pumir, D. Bensimon, V. Croquette, H.
  Richter and L. Kramer, Physica D, {\bf 55}, 269 (1992).

\bibitem{h1} F. Daviaud and J. M. Vince, 
Phys. Rev. E {\bf 48}, 4432 (1993).

\bibitem{h2} N. Mukolobwiez, A. Chiffaudel and F. Daviaud, 
Phys. Rev. Lett. {\bf 80}, 4661 (1998).

\bibitem{garnier} N. Garnier, A. Chiffaudel and F. Daviaud, unpublished results. 

\bibitem{CIAI4} N. Garnier and A. Chiffaudel, 
Phys. Rev. Lett. {\bf 86}, 75 (2001).

\bibitem{wire} J. M. Vince and M. Dubois, Physica D, {\bf 102}, 93
(1997).

\bibitem{sidewall} Y. Liu and R. E. Ecke, Phys. Rev. Lett. {\bf 78},
4391 (1997); Phys. Rev. E {\bf 59}, 4091 (1999).

\bibitem{dean} I. Mutabazi, J. J. Hegesth, C. D. Andereck and J. E. Wesfreid,
Phys. Rev. Lett. {\bf 64}, 1729 (1990);
P. Bot, O. Cadot and I. Mutabazi, Phys. Rev. E {\bf 58}, 3089 (1998);
P. Bot and I. Mutabazi, Eur. Phys. J. B {\bf 13}, 141 (2000).

\bibitem{velarde} 
A. Wierschem, H. Linde and M. G. Velarde, Phys. Rev. E {\bf 62}, 6522
(2000).

\bibitem{chem} Q. Ouyang and J. M. Flesselles, Nature {\bf 379}, 143
(1996); Q. Ouyang, H. L. Swinney and G. Li, Phys. Rev. Lett. {\bf 84},
1047 (2000); L. Q. Zhou and Q. Ouyang, Phys. Rev. Lett. {\bf 85}, 1650
(2000).

\bibitem{review} A recent and detailed review on the CGLE is
the following : I.S. Aranson and L. Kramer, {\em The World
of the Complex Ginzburg-Landau Equation}, to appear in Rev. Mod. Phys.
(cond-mat/0106115).

\bibitem{EckDiP} J. T. Stuart and R. C. DiPrima, 
Proc. R. Soc. London A {\bf 362}, 27 (1978).

\bibitem{notevHB}
The perturbation with wavenumber $k_c$ of a plane wave $q$ 
travels with a velocity $v_c = -\mbox{Im}[\sigma(k_c)]/k_c$.
At the Eckhaus instability, in the infinite system, 
$k_c\to 0$ and $v_c$ equals the group velocity 
$v_g = \partial \omega_q /\partial q = 2 q (c_1+c_3)$.
For finite system size $k_c > 0$ and $v_c \ne v_g$.

\bibitem{EckCI1} I. S. Aranson, L. Aranson, L. Kramer and A. Weber, 
Phys. Rev. A {\bf 46}, R2992 (1992).

\bibitem{EckCI2} A. Weber, L. Kramer, I. S. Aranson and L. Aranson, 
Physica D {\bf 61}, 279 (1992).

\bibitem{pumir}  A. Pumir, B. I. Shraiman, W. van Saarloos,
P. C. Hohenberg,  H. Chat\'e and M. Holen, p. 173 in
C. D. Andereck and F. Hayot (Eds.),``Ordered and Turbulent patterns in
Taylor-Couette Flow'' (Plenum Press, New York, 1992)

\bibitem{miguel} R. Montagne, E. Hern\'andez-Garc\'{\i}a and M. San Miguel,
Phys. Rev. Lett. {\bf 77}, 267 (1996); R. Montagne,
E. Hern\'andez-Garc\'{\i}a, A. Amengual and M. San Miguel, Phys. Rev. E {\bf
55}, 151 (1997).

\bibitem{torcini} A. Torcini, Phys. Rev. Lett. {\bf 77}, 1047 (1996);
 A. Torcini, H. Frauenkron and P. Grassberger, Phys.
  Rev E {\bf 55}, 5073 (1997).

\bibitem{hager} G. Hager, {\it Quasiperiodische L\"osungen der eindimensionalen
komplexen Ginzburg-Landau Gleichung}, Diploma Thesis, University of Bayreuth,
Germany (1996).

\bibitem{chao1}  B. I. Shraiman,  A. Pumir,  W. van Saarloos,
P. C. Hohenberg,  H. Chat\'e and M. Holen, Physica D {\bf 57}, 241
(1992). 

\bibitem{chate} H. Chat\'e, Nonlinearity {\bf 7}, 185 (1994); p. 33 in
 P. E. Cladis and Palffy-Muhoray (Eds.),``Spatio-Temporal Pattern Formation in
 Nonequilibrium Complex Systems'' (Addison Wesley, Reading, 1995).

\bibitem{saka} H. Sakaguchi, Prog. Theor. Phys. {\bf 84}, 792 (1990).

\bibitem{egolf} D. A. Egolf and H. S. Greenside, 
 Phys. Rev. Lett. {\bf 74}, 1751 (1995).

\bibitem{MAW1} L. Brusch, M. G. Zimmermann, M. van Hecke, M. B\"ar and
A.  Torcini, Phys. Rev. Lett. {\bf85}, 86 (2000).

\bibitem{MAW2} L. Brusch, A.  Torcini, M. van Hecke, M. G. Zimmermann
and M. B\"ar, to appear in Physica D (nlin.CD/0104029).

\bibitem{saar} W. van Saarloos and P. C. Hohenberg, Physica D {\bf
56}, 303 (1992); {\bf 69}, 209 (1993) [Errata].

\bibitem{mvh} M. van Hecke, Phys. Rev. Lett. {\bf 80}, 1896 (1998).

\bibitem{mm} M. van Hecke and M. Howard, Phys. Rev. Lett. {\bf 86},
2018 (2001).

\bibitem{note1} By substituting $\kappa\!:=\!a_z/a$ one reproduces the form of
the ODEs used in \cite{saar} which is more appropriate for studies of fronts.

\bibitem{auto97} E. Doedel, A. Champneys, T. Fairgrieve, Y. 
Kusnetsov, B. Sandstede and X. Wang, 
{\sc AUTO97}:{\it Continuation and bifurcation software for ordinary
differential equations} (Concordia University, Montreal, 1997).

\bibitem{DwightDP}
M. Kness, L. Tuckerman and D. Barkley, Phys. Rev. A {\bf 46}, 
5054 (1992). 

\bibitem{bloch} 
N. W. Ashcroft and N. D. Mermin, {\it Solid State Physics} (Holt, Rinehart and
Winston, New York, 1976);
P. Collet and J.-P. Eckmann, {\it Instabilities and Fronts in Extended Systems}
(Princeton University Press, 1990).

\bibitem{michal} M. Or-Guil, I. G. Kevrekidis and M. B\"ar, Physica D {\bf
135}, 154 (2000).

\bibitem{mvhexp} M. van Hecke, submitted to Physica D (cond-mat/0110068).

\bibitem{janiaud2} B. Janiaud, E. Guyon, D. Bensimon, and V. Croquette, 
in F. Busse and L. Kramer (Eds.),
``Nonlinear Evolution of Spatio-Temporal Structures in Dissipative Continuous
Systems'', NATO ASI Series B, Vol.225 (Plenum, New York, 1990).

\end{thebibliography}
\end{document}